%% file: Phase-Ordering_Main.tex
\documentclass[conference]{IEEEtran}
\ifCLASSINFOpdf
   \usepackage[pdftex]{graphicx}
   \usepackage{booktabs}
   \usepackage{tabularx}
   \usepackage{amsmath}
   \usepackage{xcolor, colortbl}
   \usepackage{fancyhdr}
   \usepackage{subfig}
   
\definecolor{olivegreen}{RGB}{0,128,128}
\definecolor{pink}{rgb}{1.0, 0.13, 0.32}

\usepackage{hyperref}

\newcommand{\fixme}[2][]{\textcolor{red}{\textbf{FIXME:}\ifthenelse{\isempty{#1}{}}{}{#1--}#2}}

\newcommand{\fixmecleanuppl}[2][]{\fixme{CLEAN THIS UP PLEASE}}

\newcommand\redsout{\bgroup\markoverwith{\textcolor{orange}{\rule[0.5ex]{2pt}{1pt}}}\ULon}

\else
\fi

\begin{document}
\title{POSET-RL: Phase ordering for Optimizing Size and Execution Time using Reinforcement Learning}

\author{\IEEEauthorblockN{Shalini Jain\IEEEauthorrefmark{1},
Yashas Andaluri\IEEEauthorrefmark{2}, S. VenkataKeerthy\IEEEauthorrefmark{3} and
Ramakrishna Upadrasta\IEEEauthorrefmark{4}}

\IEEEauthorblockA{Department of Computer Science and Engineering, IIT Hyderabad\\
Email: \{\IEEEauthorrefmark{1}cs15resch11010, 
\IEEEauthorrefmark{2}cs17b21m000001, 
\IEEEauthorrefmark{3}cs17m20p100001\}@iith.ac.in, 
\IEEEauthorrefmark{4}ramakrishna@cse.iith.ac.in}}

\maketitle

\begin{abstract}
\input{abstract}
\end{abstract}

\IEEEpeerreviewmaketitle

\input{introduction}
\input{background}

\input{methodology}
\input{selectionOfSubSequences}
\input{experimentationAndResultAnalysis}
\input{relatedWorks}
\input{conclusion}

\section*{Acknowledgments}
We are grateful to Suresh Purini for insightful discussions and guidance at the early stage of this work. We thank Anilava Kundu for the discussions and help in the initial stages of implementation.

This research is funded by the Department of Electronics \& Information Technology and the Ministry of Communications \& Information Technology, Government of India. This work is partially supported by a Visvesvaraya PhD Scheme under the MEITY, GoI (PhD-MLA/04(02)/2015-16), an NSM research grant (MeitY/R$\&$D/HPC/2(1)/2014), a Visvesvaraya Young Faculty Research Fellowship from MeitY, and a Google PhD Fellowship.

\bibliographystyle{ieeetr}
{\bibliography{references}}

\end{document}

%% file: abstract.tex
The ever increasing memory requirements of several applications has led to increased demands which might not be met by embedded devices. 
Constraining the usage of memory in such cases is of paramount importance.
It is important that such code size improvements should not have a negative impact on the runtime.
Improving the execution time while optimizing for code size is a non-trivial but a significant task.

The ordering of standard optimization sequences in modern compilers is fixed, and are heuristically created by the compiler domain experts based on their expertise.
However, this ordering is sub-optimal, and does not generalize well across all the cases.

We present a reinforcement learning based solution to the phase ordering problem, where the ordering improves both the execution time and code size. 
We propose two different approaches to model the sequences: one by manual ordering, and other based on a graph called \textit{Oz Dependence Graph (ODG)}.
Our approach uses minimal data as training set, and is integrated with LLVM.

We show results on x86 and AArch64 architectures on the benchmarks from SPEC-CPU 2006, SPEC-CPU 2017 and MiBench.
We observe that the proposed model based on ODG outperforms the current \texttt{Oz} sequence both in terms of size and execution time by 6.19\% and 11.99\% in SPEC 2017 benchmarks, on an average.

\textit{\textbf{Keywords-}}{Phase Ordering, Compiler Optimization, Reinforcement learning}

%% file: introduction.tex
\section{Introduction}
One of the essential components of compiler construction are optimizations. They involve transforming the input program and generating code that is semantically equivalent to it, but performs better in various metrics like execution time, code size, power utilization, etc.

Modern compilers have several number of  optimizations like vectorization, dead code elimination implemented as passes.
A collection of these individual passes are statically ordered \textit{a priori} and are exposed as \textit{off-the-shelf} standard optimization flags like \texttt{O0, O1, O2, O3, Os,} and \texttt{Oz}.
Among these predefined optimization flags, \texttt{O0} disables all optimizations, essentially making the compiler a translator of the source-program to binary. It provides fast compilation of source code, while also giving good correlation between source and the generated code. 
Optimization flags \texttt{O1, O2,} and \texttt{O3} are designed to improve performance by reducing the execution time. While the \texttt{O1} flag enables some important optimizations, the \texttt{O2} flag adds a few more optimizations.  
The \texttt{O2} flag also changes the heuristics for the optimizations when compared to \texttt{O1} flag.
With respect to execution of the code, the \texttt{O3} sequence gives the best performance among all these optimizations. 
Although these optimizations improve the performance of program execution, they may also increase code size.

The \texttt{Os} and \texttt{Oz} flags are designed to optimize for code size. The \texttt{Os} flag is designed to perform nearly similar to \texttt{O2} flag, while promising an appreciable decrease in code size.
The \texttt{Oz} flag is designed to provide more code size reduction than \texttt{Os}.
Such a reduction in code size by optimizations in \texttt{Oz}, generally incurs penalty on the execution time, leading to a trade-off between \texttt{O3} and \texttt{Oz} in terms of code size and execution time. 

\begin{figure}
    \centering
    \includegraphics[scale=0.55]{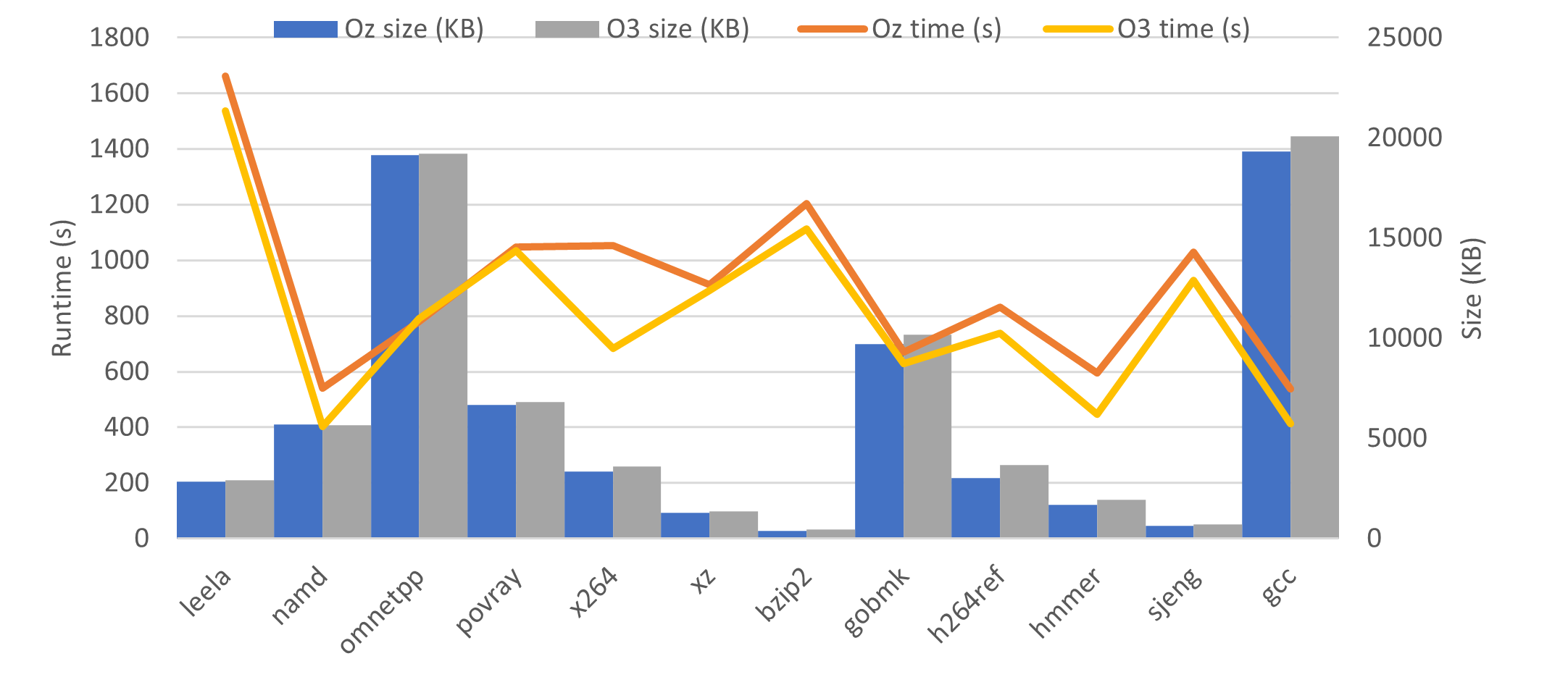}
    \caption{\texttt{O3} vs. \texttt{Oz}: Comparison of runtime and code size}
    \label{fig:O3Oz}
\end{figure}

We observe on a set of benchmarks from the standard SPEC CPU benchmarks that the binaries generated by \texttt{Oz} flag incurs $10\%$ more execution time than that of \texttt{O3}; while resulting in about $3.5\%$ improvement in terms of code size, on an average. A chart showing the runtime and code size characteristics of these benchmarks is shown in Fig.~\ref{fig:O3Oz}. In this work, we try to find if this gap exhibited by \texttt{Oz} passes can be improved by further reducing the code size while improving the run time.

Although these predefined optimization flags achieve good performance when compared to the unoptimized case, they do not guarantee the best performance for every given input program. Moreover, the ordering of passes in these sequences is not based on a particular input program but is statically predefined. 

The problem of selecting and ordering the optimization passes, called the \textit{phase ordering problem} has been a fertile ground for compiler researchers for many decades~\cite{PhaseCoupling}. 
The classic paper by Cooper et al.~\cite{Cooper99} refers to this particular problem where they propose using AI/ML techniques for reordering optimizations. 

In the literature, the phase ordering problem has been recognized to be an important problem for several reasons, including the following:
\begin{itemize}

    \item On a particular program, two optimizations applied in different orders may have an unpredictable impact on its performance. It is also possible that a predetermined optimization sequence could result in an adverse effect on the performance of the program.
    
    \item After transformation of a program by an optimization sequence or flag, there is no guarantee to ensure performance improvements. The user, who is interested in the performance improvement of \textit{a particular program}, may also expect some guarantees about the potential improvements on the particular program, or even a assurance that all (possible) optimizations have been tried to improve the program.

    \item The \textit{best optimization order} designed for one program may not perform well on another (new) program. The predetermined optimization sequence could even degrade the performance for new programs.
    
\end{itemize}

Exploiting the undecidable nature of this problem~\cite{Touati}, several machine learning based approaches~\cite{MiCOMP},~\cite{Fursin2008MILEPOSTGM},~\cite{Mammadli},~\cite{PuriniS},~\cite{YACOS} have been used to predict the optimization sequences given an input program. These techniques use the expert-designed \textit{program features} to represent programs for training the model to arrive at the solution. 

Several approaches have been proposed in the recent times to represent programs as vectors, called \textit{program embeddings}, analogous to word2vec like approaches of natural language processing.
Such embeddings created by considering syntactic and semantic information of the program is used as the input for ML purposes.
The most notable embeddings are AST based embeddings of code2vec~\cite{alon2019code2vec}, and the IR based embeddings of inst2vec~\cite{ncc}, and IR2Vec~\cite{IR2Vec}. 

To our understanding, the earlier works on phase ordering have considered code size and execution time optimizations as \textit{independent} objectives.
However, optimizing for codesize and execution time are orthogonal - optimization for codesize can adversely affect the execution time and vice versa.
On the other hand, if the execution time is not considered while optimizing for codesize, gains in codesize can lead to worse performance. 
Hence, we propose a Reinforcement Learning (RL) based solution for determining better optimization sequences for a given program to optimize for \textit{both} code size together and the execution time in LLVM compiler.
Our work is based on modeling programs as vectors using IR2Vec representations which is a LLVM-based program encoding.

We use a reinforcement learning approach as the search space of optimizations is too big to enumerate.
For a compiler with $m$ optimization passes, if the sequence length is fixed as $n$, then there can be potentially $m^n$ combinations, allowing repetitions.
And, \texttt{Oz} of LLVM has 90 transformation passes, among which 54 are unique.
If one has to replicate such a setup, there can be potentially $54^{90}\approx 10^{156}$ combinations.
As it can be seen, creating this many combinations for each element of the dataset is highly infeasible, making reinforcement learning a go to method for this problem.

Also, we use IR2Vec representations as it captures the syntax and semantics of the code intricately when compared to the expert-designed feature set.
IR2Vec representations of the program are modelled as the state, and the sequence of optimization passes as actions for the formulated RL problem. The agent predicts the optimizations to apply on the input program. The environment - LLVM optimizer applies the predicted sequences and recomputes the representation corresponding to the predicted sequence to update the state for the next step. This environment is thus responsible for handling the interaction between the agent and the effect of the action it takes in the form of rewards. 

We propose two different sets of sub-sequences to model the action space in our RL model. 
Similar to the earlier works like~\cite{MiCOMP}, \cite{Mammadli} that model optimization sequences for execution time by manual or analytical grouping of passes, we too design a logical grouping of optimization passes to form the action space. These groups form the sub-sequences, and are obtained from the standard \texttt{-Oz} sequence of LLVM. 
Second, we create a graph: \texttt{Oz} Dependence Graph (ODG) depicting the dependencies among the passes in the \texttt{Oz} sequence and derive the sequences by traversing the graph. 

Our method uses Q-learning~\cite{watkins1992q}, which involves determining the optimum Q-function for a state-action pair. We approximate this by using a neural network called Deep Q-Network~\cite{mnih2013playing},\cite{mnih2015human}, which performs better with complex environments than the simple Q-learning algorithm that maintains a table or a matrix called Q-table. We trained our model with the single source llvm-test-suite benchmark files~\cite{llvm-test-suite}. 
Our trained model is able to reduce binary sizes for most of the benchmarks from MiBench~\cite{mibench}, SPEC CPU (int and fp) 2017~\cite{spec17-Bucek:2018:SCN:3185768.3185771}, and SPEC CPU 2006~\cite{spec06:journals/sigarch/Henning06} benchmark-suites. 

\textbf{Contributions:}
The following are our key contributions:
\begin{itemize}
    \item We propose a novel Deep Q-Network (DQN) based RL model that predicts the optimal sequence of optimization passes for a given input source program. Our framework is unique as it builds from embeddings given by IR2Vec encoding and derives its scalability characteristics.
   
    \item We propose a \textit{\texttt{Oz} Dependence Graph (ODG)} to cluster the individual optimization passes into various sub-sequences that forms an action space of the RL model. ODG shows all the dependencies among the optimization passes in terms of the order they are present in the \texttt{Oz} flag.
    
    \item We study the results on two different architectures: x86 and AArch64. We show results on MiBench, SPEC CPU 2006 and 2017 benchmark suites.
    
    \item We show that our solution outperforms the standard \texttt{Oz} sequence by achieving better code size reduction while improving the runtime on standard compiler benchmarks. In SPEC 2017, 2006 and MiBench, we achieve a maximum size reduction of upto $22.94\%$, $9.93\%$, $8.68\%$, while achieving improvements in runtimes of upto $46\%$, $6\%$ and $16\%$ respectively.
\end{itemize}

The rest of the paper is further organized as the following: Some relevant background of Reinforcement Learning (RL) is explained in Section~\ref{sec:background}. In our solution, we model each program as states, and optimization passes as actions in accordance with reinforcement learning techniques. 
We discuss this design in Section~\ref{sec:methodology}. 
In Section~\ref{sec:sub-sequences}, we explain the rationale for the above design. 
In Section~\ref{sec:experimentation}, we discuss experimentation details, and the detailed analysis of results.  
We discuss the related works in Section~\ref{sec:relatedworks}. 
Finally, in Section~\ref{sec:conclusion}, we conclude our work. 

%% file: background.tex
\section{Background}

\label{sec:background}
In this section we give a brief summary of the reinforcement learning techniques and the program representation method that we use.

\begin{figure}
    \centering
    \includegraphics[scale=0.4]{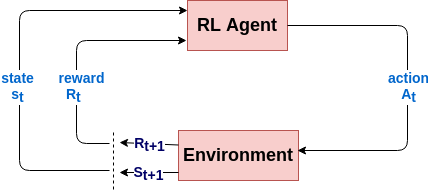}
    \caption{Reinforcement Learning Process}
    \vspace{-\baselineskip}
    \label{fig:RLflow}
\end{figure}

\subsection{Reinforcement Learning and Markov Decision Process} 
The overview of a reinforcement learning method using Markov Decision Process (MDP) is shown in Fig.~\ref{fig:RLflow}. A reinforcement learning problem modelled on the Markov Decision Process usually consists of the following: 

\begin{itemize}
    \item \textbf{Agent:} An agent is the actor that makes a decision by choosing an action from an action space for which it receives reward from the environment. A positive valued reward represents a good decision and negative value represents a bad decision. After performing any action, the agent moves from current state to the new state.
    \item \textbf{States:} An instance of the environment is defined as a state. For each action the agent takes, the current state transitions to a new state. 
    \item \textbf{Action:} Decision taken by the agent by looking into the current state. Set of all actions that an agent can take forms an \textit{action space}.
    \item \textbf{Environment:} The agent is placed in an environment that rewards or punishes the agent for the current action. The agent interacts with the environment to learn better so as to maximize the reward in future.

\end{itemize}

 An agent usually interacts with the environment through states and actions to get rewards.
  The formulation of RL problems can be exploited by a software agent to explore an environment and make profitable decisions. This fits nicely with our need for exploiting a sequence of actions viz optimization passes and taking profitable actions that results in optimizing code size along with the execution time.

 As discussed earlier, reinforcement learning can be formulated using Markov Decision Processes (MDPs). An MDP provides a mathematical framework that can model decision making. An agent is said to follow the Markov Property, if the current state of the agent depends only on its immediate previous state or its next state depends only on its current state.
 This augurs well with respect to our problem where the agent can choose an optimization pass given its current state and not have to look in the past.

\subsection{Deep Q-Networks}
Q-learning involves quantifying state-action pairs with q-values that define how good a given action for a particular state is. 
The vanilla Q-learning in practice involves maintaining a table or matrix of values. 
Although the Q-learning algorithm may generalize well to real-world problems,
it usually becomes infeasible for complex environments as the number of states go higher.
The Deep Q-Network (DQN) was introduced to solve this problem which involves combining the traditional Q-learning algorithm with Deep Neural Networks~\cite{mnih2013playing}. 
The deep neural network here acts as a function approximator for the Q-function by replacing the need for storing values in a Q-table. 
The various flavours of DQN have been proposed and are widely used~\cite{mnih2013playing},~\cite{mnih2015human},~\cite{van2016deep},~\cite{wang2016dueling}. 
We use Double DQN for modelling our problem. 

A vanilla DQN on training suffers from the overestimation of Q values~\cite{van2016deep}. 
The notion of double DQN was introduced to meet with this challenge. 
Instead of using the target network for predicting the future q-value, double DQN uses an online network to predict the best action $a'$ for a given state $s'$. 
The q-value, $q(s',a')$ for the state-action pair $(s',a')$ is given by the target network.

\subsection{IR2Vec}

As described earlier, modern compilers have several nontrivial compiler optimizations implemented in them that are the result of manual heuristics designed by expert compiler writers. Various machine learning techniques have been proposed to solve such tasks on source code by modelling them as Natural Language Processing problems~\cite{allamanis2018survey}.
Of late, researchers identified the issues arising from modelling programs as natural languages and proposed programming language centric approaches to model programs.
These approaches result in representing programs as vectors~\cite{alon2019code2vec},~\cite{ncc},~\cite{IR2Vec}, analogous to that of natural language representations like word2vec and gloVe.

We use IR2Vec representations~\cite{IR2Vec} to represent programs in our model.
IR2Vec uses the intermediate representation of LLVM to represent programs as high dimensional vectors called program embeddings.
As it is based on LLVM IR, the representation is naturally independent of the source languages and target architectures. It can generate encodings for all the languages that LLVM can support. 

The IR2Vec framework is primarily built from a vocabulary containing the representation of each fundamental entities of the IR, viz. opcode, type and operands. These are used to form the higher-level representations, at the program level, function level, or instruction level.  Also, IR2Vec uses the \textit{use-def}, \textit{reaching definitions}, and \textit{live variable} analyses information to construct flow-analysis based representations. It is shown that IR2Vec representations have high scalablity, have better memory characteristics, and exhibit better Out Of Vocabulary characteristics than the other approaches.

We chose IR2Vec representations to represent programs, as it builds on program analysis approaches to form program centric representations.

%% file: methodology.tex
\section{Methodology}
\label{sec:methodology}
\begin{figure}
    \centering
    \includegraphics[scale=0.31]{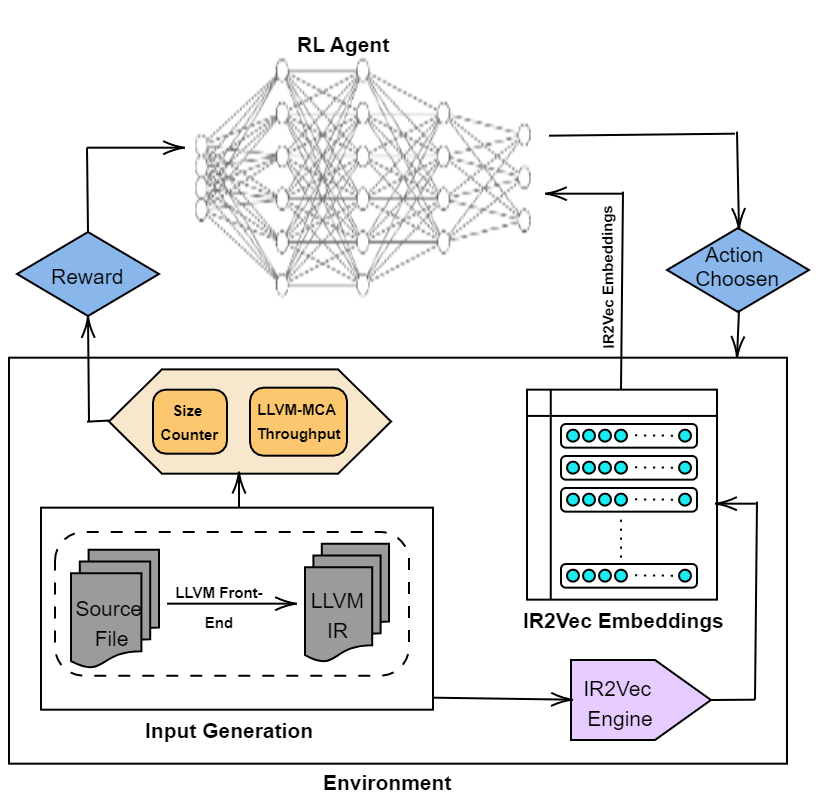}
    \caption{Overview of the proposed workflow}
    \vspace{-\baselineskip}
    \label{fig:flowgraph}
\end{figure}

The overview of our proposed methodology is shown in Fig.~\ref{fig:flowgraph}. 
The input source code is initially fed to the environment where the LLVM IR is generated by the LLVM’s front-end, which in turn is converted to embeddings by using the IR2Vec module as shown in Fig.~\ref{fig:flowgraph}. 
These representations of input programs form our observation space or states as discussed in Section~\ref{sec:background}. 
The RL agent starts learning the information about the environment through a series of episodes by consuming these embeddings as input.
For each step in an episode, given the input program representation, the agent interacts with the environment by predicting the optimization sub-sequence to apply as an action from the action space. We describe the action space in detail in Section~\ref{sec:sub-sequences}.
The LLVM optimizer or opt is used to apply the optimizations predicted by the RL agent, which then transitions the current state to the next state. 
The new state is formed by updating the IR2Vec representations corresponding to updated code after applying the predicted transformations. 
For each such transition, the RL agent receives a reward and continues the same till it reaches the end of an episode.
We refer to this flow diagrammatically in Fig.~\ref{fig:flowgraph}. 
The rewards are defined as the combination of reduction in the size of an object file as well as improvement in throughput in comparison to an object file optimized with the \texttt{Oz} sequence.

We broadly divide our framework into 3 main modules viz (i)  Environment, (ii) RL agent, and (iii) Reward computation which we explain in greater detail in subsequent Sections,~\ref{subsec:environment},~\ref{subsec:agent} and~\ref{subsec:rewards} respectively. 

\subsection{Environment}
\label{subsec:environment}
As our work is based on LLVM-IR, we convert source files into LLVM-IR using the existing  frontend framework. Since we consider only C/C++ input files, clang/clang++ converts it into IR files. These IR files are  original non-optimized versions. 
We consider this data as an input dataset to the RL agent.

As discussed in Section~\ref{sec:background}, the environment in an RL problem allows the agent to explore and make decisions on the basis of the actions it takes. Here our environment serves the same for the RL agent which interacts with it to perform actions and to get rewards associated with it. Our environment further consists of three modules i.e. IR2Vec embeddings, Rewards and action space, each of these discussed in detail below.

\subsubsection{IR2Vec Embeddings}
To represent the source programs we use IR2Vec~\cite{IR2Vec} embeddings which represent each program as higher dimensional vectors that encode both program features as well as flow information. This enables us to better represent the source programs as the semantics of it is also encoded in the embeddings. Each source program vector acts as a state for the environment which changes with the application of optimization passes. This causes the  DQN to have transitioned to a different state.

\subsubsection{Rewards}
During training, we compile the LLVM-IR module obtained at each step to its object file, without linking, so as to prevent the inclusion of code from linked libraries. 
This module computes the total size of the object file.
We use LLVM-MCA~\cite{llvm-mca} to estimate the throughput of the generated binary.
Higher the throughput, lesser would be the runtime.
Hence we use this throughput as a static measure to model the runtime of the code.
Both the code size and throughput components form the reward.
We describe reward calculation in section~\ref{subsec:rewards}. 
Initially, for each source program in the input dataset, this module computes the size of an object file and throughput generated by LLVM-MCA without any optimization and with the \texttt{-Oz} optimization sequence. 
Then, for each step of an episode, this module computes the size of an object file and throughput generated by LLVM-MCA for the resultant object file on applying the predicted action.

\subsubsection{Action Space}
Action space consists of all the sub-sequences (set of optimizations passes from LLVM) generated from the \texttt{Oz} optimization sequence. At each step the RL agent selects any one sub-sequence from the action space and applies it to the IR present in the current state. We experiment with two approaches for forming the action space - sub-sequences generated by manual grouping and sub-sequence selection by ODG, discussed in detail in Section~\ref{sec:sub-sequences}.

\subsection{RL Agent} 
\label{subsec:agent}
As discussed in Section~\ref{sec:background}, we define our RL agent as a Double Deep Q-Network (DDQN).
Given the IR2Vec representation as a state, the DDQN predicts the associated Q-value for every action in the action space containing different optimization sequences.
As discussed earlier, the optimization sequence with the highest Q-value is chosen and is then applied to LLVM-IR of the source program to obtain the next state.
After gathering enough experiences by interacting with the environment, the DDQN then starts to train to obtain the optimal Q-value for every state.

\subsection{Reward Computation}
\label{subsec:rewards} 
We define our reward (R) as the combination of improvement in binary size and the throughput corresponding to the optimized and the original, non optimized versions of the compiled object file. It is computed as:
\begin{equation}
    R = \alpha * R_{BinSize} + \beta * R_{Throughput}
    \label{eqn:Reward}
\end{equation}
Where, reward corresponding to BinSize ($R_{BinSize}$) is defined as the ratio of change in size between each interim states of an episode to binary size of the original object file, as shown in Eqn~(\ref{eqn:BinSize}).
\begin{equation}
    R_{BinSize} = \frac{BinSize_{last} - BinSize_{curr}}{BinSize_{base}}
    \label{eqn:BinSize}
\end{equation}
Here, $BinSize_{last}$ refers to the size of the object file of the previous state, $BinSize_{curr}$ refers to the total size of the object file after performing the action chosen by the model, and $BinSize_{base}$ refers to the size of the object file without any optimization.

We describe the throughput component of the reward, $R_{Throughput}$ as the ratio of change in throughput between each interim states of an episode to the throughput of the original object file, as shown in Eqn~(\ref{eqn:Throughput}). 
\begin{equation}
    R_{Throughput} = \frac{Throughtput_{curr} - Throughtput_{last}}{Throughtput_{base}}
    \label{eqn:Throughput}
\end{equation}

Here, $Throughput_{curr}$, $Throughput_{last}$ and $Throughput_{base}$ refer to the throughput generated by LLVM-MCA for current state, last state and without any optimization respectively.

%% file: selectionOfSubSequences.tex
\section{Selection of Sub-Sequences}
\label{sec:sub-sequences}

\begin{table*}[ht]
    \centering
    \caption{List of transformation passes in \texttt{Oz} optimization sequence (LLVM-10)}
    \begin{tabularx}{\textwidth}{p{0.97\linewidth}}
         \toprule
         \textbf{List of Transformation Passes in \texttt{Oz}} \\
         \midrule
         -ee-instrument -simplifycfg -sroa -early-cse -lower-expect -forceattrs -inferattrs -ipsccp -called-value-propagation -attributor -globalopt -mem2reg -deadargelim -instcombine -simplifycfg -prune-eh -inline -functionattrs -sroa -early-cse-memssa -speculative-execution -jump-threading -correlated-propagation -simplifycfg -instcombine -tailcallelim -simplifycfg -reassociate -loop-simplify -lcssa -loop-rotate -licm -loop-unswitch -simplifycfg -instcombine -loop-simplify -lcssa -indvars -loop-idiom -loop-deletion -loop-unroll -mldst-motion -gvn -memcpyopt -sccp -bdce -instcombine -jump-threading -correlated-propagation -dse -loop-simplify -lcssa -licm -adce -simplifycfg -instcombine -barrier -elim-avail-extern -rpo-functionattrs -globalopt -globaldce -float2int -lower-constant-intrinsics -loop-simplify -lcssa -loop-rotate -loop-distribute -loop-vectorize -loop-simplify -loop-load-elim -instcombine -simplifycfg -instcombine -loop-simplify -lcssa -loop-unroll -instcombine -loop-simplify -lcssa -licm -alignmentfromassumptions -strip-dead-prototypes -globaldce -constmerge -loop-simplify -lcssa -loop-sink -instsimplify -div-rem-pairs -simplifycfg \\
         \bottomrule
    \end{tabularx}
    \label{tab:Oz_Sequence}
\end{table*}

{
\begin{table*}[h]
    \centering
    \caption{List of manual sub-sequences designed from \texttt{Oz} sequence}
    \begin{tabularx}{\textwidth}{r|l}
        \toprule
        S.No. & Manual Sub-sequence \\
        \midrule
        1 & -ee-instrument -simplifycfg -sroa -early-cse -lower-expect -forceattrs -inferattrs -mem2reg \\
        2 & -ipsccp -called-value-propagation -attributor -globalopt \\
        3 & -deadargelim -instcombine -simplifycfg \\
        4 & -prune-eh -inline -functionattrs -barrier \\
        5 & -sroa -early-cse-memssa -speculative-execution -jump-threading -correlated-propagation \\
        6 & -simplifycfg -instcombine -tailcallelim -simplifycfg -reassociate \\
        7 & -loop-simplify -lcssa -loop-rotate -licm -loop-unswitch -simplifycfg -instcombine \\
        8 & -loop-simplify -lcssa -indvars -loop-idiom -loop-deletion -loop-unroll \\
        9 & -mldst-motion -gvn -memcpyopt -sccp -bdce -instcombine -jump-threading -correlated-propagation -dse \\
        10 & -loop-simplify -lcssa -licm -adce -simplifycfg -instcombine \\
        11 & -barrier -elim-avail-extern -rpo-functionattrs -globalopt -globaldce -float2int -lower-constant-intrinsics \\
        12 & -loop-simplify -lcssa -loop-rotate -loop-distribute -loop-vectorize \\
        13 & -loop-simplify -loop-load-elim -instcombine -simplifycfg -instcombine \\
        14 & -loop-simplify -lcssa -loop-unroll -instcombine -loop-simplify -lcssa -licm -alignment-from-assumptions \\
        15 & -strip-dead-prototypes -globaldce -constmerge -loop-simplify -lcssa -loop-sink -instsimplify -div-rem-pairs -simplifycfg \\
        \bottomrule
    \end{tabularx}
    \label{tab:manual_subsequences}
\end{table*}
}

\begin{figure*}
    \centering
    \includegraphics[scale=0.43]{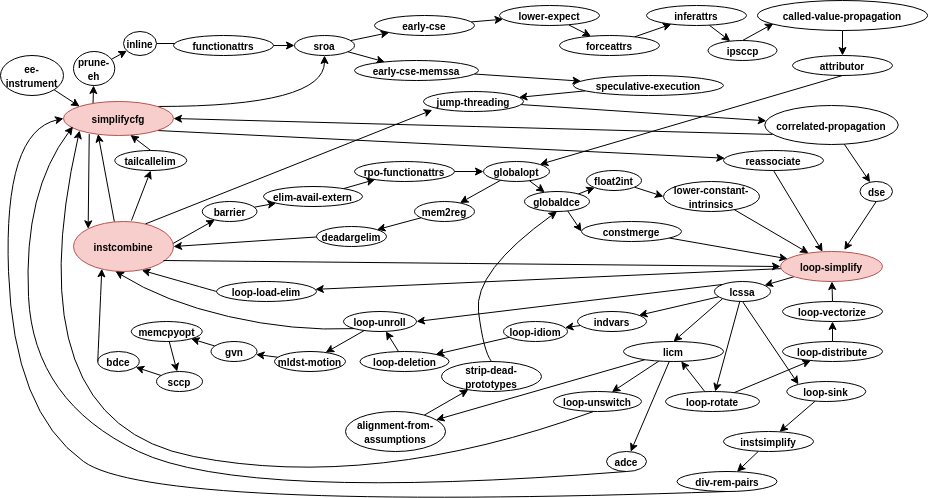}
    \caption{\texttt{Oz} Dependence Graph}
    \vspace{-\baselineskip}
    \label{fig:ODG}
\end{figure*}

\begin{table*}[h]
    \centering
    \caption{List of sub-sequences designed from ODG of \texttt{Oz} sequence}
    \begin{tabularx}{\textwidth}{r|l}
        \toprule
        S.No. & ODG Sub-sequence \\
        \midrule
        1 & -instcombine -barrier -elim-avail-extern -rpo-functionattrs -globalopt -globaldce -constmerge \\
        2 & -instcombine -barrier -elim-avail-extern -rpo-functionattrs -globalopt -globaldce -float2int -lower-constant-intrinsics \\
        3 & -instcombine -barrier -elim-avail-extern -rpo-functionattrs -globalopt -mem2reg -deadargelim \\
        4 & -instcombine -jump-threading -correlated-propagation -dse \\
        5 & -instcombine -jump-threading -correlated-propagation \\
        6 & -instcombine \\
        7 & -instcombine -tailcallelim \\
        8 & -loop-simplify -lcssa -indvars -loop-idiom -loop-deletion -loop-unroll \\
        9 & -loop-simplify -lcssa -indvars -loop-idiom -loop-deletion -loop-unroll -mldst-motion -gvn -memcpyopt -sccp -bdce \\
        10 & -loop-simplify -lcssa -licm -adce \\
        11 & -loop-simplify -lcssa -licm -alignmentfromassumptions -strip-dead-prototypes -globaldce -constmerge \\
        12 & -loop-simplify -lcssa -licm -alignmentfromassumptions -strip-dead-prototypes -globaldce -float2int -lower-constant-intrinsics \\
        13 & -loop-simplify -lcssa -licm -loop-unswitch \\
        14 & -loop-simplify -lcssa -loop-rotate -licm -adce \\
        15 & -loop-simplify -lcssa -loop-rotate -licm -alignmentfromassumptions -strip-dead-prototypes -globaldce -constmerge \\
        16 & -loop-simplify -lcssa -loop-rotate -licm -alignmentfromassumptions -strip-dead-prototypes -globaldce -float2int -lower-constant-intrinsics \\
        17 & -loop-simplify -lcssa -loop-rotate -licm -loop-unswitch \\
        18 & -loop-simplify -lcssa -loop-rotate -loop-distribute -loop-vectorize \\
        19 & -loop-simplify -lcssa -loop-sink -instsimplify -div-rem-pairs -simplifycfg \\
        20 & -loop-simplify -lcssa -loop-unroll \\
        21 & -loop-simplify -lcssa -loop-unroll -mldst-motion -gvn -memcpyopt -sccp -bdce \\
        22 & -loop-simplify -loop-load-elim \\
        23 & -simplifycfg \\
        24 & -simplifycfg -prune-eh -inline -functionattrs -sroa -early-cse -lower-expect -forceattrs -inferattrs -ipsccp -called-value-propagation -attributor\\
        & -globalopt -globaldce -constmerge -barrier \\
        25 & -simplifycfg -prune-eh -inline -functionattrs -sroa -early-cse -lower-expect -forceattrs -inferattrs -ipsccp -called-value-propagation -attributor\\
        & -globalopt -globaldce -float2int -lower-constant-intrinsics -barrier \\
        26 & -simplifycfg -prune-eh -inline -functionattrs -sroa -early-cse -lower-expect -forceattrs -inferattrs -ipsccp -called-value-propagation -attributor\\
        & -globalopt -mem2reg -deadargelim -barrier \\
        27 & -simplifycfg -prune-eh -inline -functionattrs -sroa -early-cse-memssa -speculative-execution -jump-threading -correlated-propagation -dse -barrier \\
        28 & -simplifycfg -prune-eh -inline -functionattrs -sroa -early-cse-memssa -speculative-execution -jump-threading -correlated-propagation -barrier \\
        29 & -simplifycfg -reassociate \\
        30 & -simplifycfg -sroa -early-cse -lower-expect -forceattrs -inferattrs -ipsccp -called-value-propagation -attributor -globalopt -globaldce -constmerge \\
        31 & -simplifycfg -sroa -early-cse -lower-expect -forceattrs -inferattrs -ipsccp -called-value-propagation -attributor -globalopt -globaldce -float2int\\
        & -lower-constant-intrinsics \\
        32 & -simplifycfg -sroa -early-cse -lower-expect -forceattrs -inferattrs -ipsccp -called-value-propagation -attributor -globalopt -mem2reg -deadargelim \\
        33 & -simplifycfg -sroa -early-cse-memssa -speculative-execution -jump-threading -correlated-propagation -dse \\
        34 & -simplifycfg -sroa -early-cse-memssa -speculative-execution -jump-threading -correlated-propagation \\
        \bottomrule
    \end{tabularx}
    \label{tab:ODG_subSequences}
\end{table*}

LLVM provides several optimization sequences (\texttt{O0, O1, O2, O3, Os,} and \texttt{Oz}) that contribute either or both to enhancing speedup and reduction of code size. \texttt{Os} and \texttt{Oz} optimization sequences are well known for code size reduction. The \texttt{Oz} optimization sequence aims to reduce code size of the given LLVM IR (even further than \texttt{Os}, at the cost of possibly greater execution time than \texttt{O2}/\texttt{O3}). 

The \texttt{Oz} sequence has the goal of reducing code size and is designed using the same set of individual optimization passes as \texttt{O2}. However, the individual optimizations occur with different frequencies, and in a different order when compared to \texttt{O2}. And, some passes vary the parameters for optimizations depending on the optimization level, as different levels have different goals.

Finding the optimal order of optimization passes with repetitive occurrences of optimizations has always been a very hard problem. 
One simple way to define action space is to make every individual pass as an action for our RL model. Firstly, this has potential to induce a very large search space, as it grows exponentially (due to various possible combinations of individual passes, and sequence lengths). Further, this type of modeling faces the following complications:
\begin{itemize}
    \item Certain transformation passes cannot be applied individually. They must appear in a specific order to maintain their mutual dependencies.
    \item Transformation passes can potentially disturb various properties of the IR: either the control flow of the program, or its SSA property. These properties must be restored by other passes before other transformation passes are applied on the IR.
\end{itemize}

Due to these reasons, several works on phase ordering proposed using different set of optimization passes, commonly known as sub-sequences with respect to improving the execution time~\cite{MiCOMP}, \cite{Mammadli}, \cite{PuriniS}, code size reduction~\cite{YACOS} and power utilization~\cite{pallister2015identifying}. 
Similarly, in this work we too prefer to use sub-sequences instead of individual passes as actions. 
We generate these sub-sequences by utilizing the \texttt{Oz} sequence, as \texttt{Oz} is better known for code size reduction. This reduces the search space of the model and also leads in better code size reduction.

In this work, we propose two approaches to form the action space: the first is to group the optimization passes manually to form sub-sequences.
The second approach involves forming a graph out of the optimization passes from \texttt{Oz} sequence, and traversing it to form sub-sequences, while preserving their dependencies.
These two approaches are discussed in Sec.~\ref{section4.1} and Sec.~\ref{section4.2}.

\subsection{Sub-sequences generated by manual grouping}
\label{section4.1}
As the number of passes is finite, obtaining precise sub-sequences based on their functionality can help the model make better decisions for selecting a sub-sequence. The logic behind this is, as for different programs, different sub-sequences may have different performance effects. So the  model can learn better based on the program characteristics with the help of IR2Vec embedding and make better decisions.


Table~\ref{tab:Oz_Sequence} shows the \texttt{Oz} optimization sequence for llvm-10. The sub-sequences created manually from LLVM’s \texttt{Oz} optimization sequence are listed in Table~\ref{tab:manual_subsequences}. This manual grouping of passes has been done by looking into specific functionalities of passes and the \texttt{Oz} sequence. For example, sequence 2 performs some module level optimizations, while sequence 4 performs inlining. Similarly, sequence 7 contains loop related passes like loop rotate, loop invariant code motion, loop unswitch and also combines redundant instructions, while sequence 12 performs loop distribution and vectorization. 


\subsection{Sub-sequences generated by \texttt{Oz} Directed Graph (ODG)}
\label{section4.2}

We utilize the ordering of the passes in \texttt{Oz} to create new sequences. We propose a directed graph (ODG) that is derived from the ordering of \texttt{Oz} passes, as shown in Fig.~\ref{fig:ODG}.

Each transformation pass in the \texttt{Oz} flag shown in the Table~\ref{tab:Oz_Sequence} forms the node, and the edges depict the ordering of the passes within the sequence. 
I.e., there will be an edge between two nodes if they appear consecutively in the \texttt{Oz} pass sequence. 
For example, as \texttt{simplifycfg} appears after \texttt{instcombine} in \texttt{Oz}, we create a directed edge from \texttt{simplifycfg} to \texttt{instcombine} in the graph.

We pick nodes with a degree greater than an arbitrary value $k$ as \textit{critical nodes}. 
Such critical nodes are the prominent nodes in the graph, where more than $k$ passes depend on them. Hence, we start and end the traversal on the ODG with such nodes.
ODG, when traversed from a critical node, would result in a walk where the dependencies of each pass would appear before it. We end the traversal when we reach another critical node.
We model the sub-sequences for action space using the sequence of nodes that are visited during this traversal.
This way of obtaining sequences from the walk on ODG allows us to get the sub-sequences that are not part of the \texttt{Oz} sequence while keeping the dependencies of the passes intact.

We choose a degree $k\geq8 $ for determining the critical nodes. Consequently, we have \texttt{simplifycfg}, \texttt{instcomibne} and, \texttt{loop-simplify} as critical nodes. They have a degree of 11, 10 and 8 respectively.
This results in 34 different sub-sequences in our action space is shown in Table~\ref{tab:ODG_subSequences}.

%% file: experimentationAndResultAnalysis.tex
\section{Experimentation and Result analysis}
\label{sec:experimentation}

In this section we talk about the experimental setup of our framework along with our results. 

\subsection{Experimental Setup}
\label{ex-setup}
We used LLVM-10 for the experiments on the proposed framework.
The modular structure of LLVM provides us with easy integration of our proposed action spaces viz Manual sub-sequences listed in Table~\ref{tab:manual_subsequences} and sub-sequences generated with ODG listed in Table~\ref{tab:ODG_subSequences}. 
The IR2Vec~\cite{IR2Vec} encoding and infrastructure which we choose to represent input source programs also fits nicely with our choice for LLVM-10.

We use 130 source code files from the single source benchmarks of LLVM-test-suite~\cite{llvm-test-suite}. 
And, we consider entirely different set of programs/benchmarks from MiBench, LLVM-test-suite, SPEC-CPU 2006 and SPEC-CPU-2017 benchmark suite for validation.
We use the program level embeddings of IR2Vec which represents a complete source program as a vector of 300 dimensions.

\paragraph*{Training}
The proposed DQN model starts the training in batches on the input training data. Each batch is a random sample of states of IR2Vec embeddings which is generated by the IR2Vec module to form the state space for the base environment.

The implementation of DQN follows $\epsilon$-greedy algorithm with $\epsilon$ set to a high value initially and then annealed in subsequent steps. When $\epsilon$ is set to a high value the DQN is said to be in the \textbf{exploration} state, which involves randomly selecting actions and remembering the associated rewards and the next state in a buffer called \textbf{replay memory}. The rewards for each action is calculated using the reward function R described in section~\ref{subsec:rewards}.
We set $\alpha$ to 10 and $\beta$ to 5 in the reward function to give more weight to $R_{BinSize}$ than $R_{Throughput}$. 
With the annealing of $\epsilon$, the DQN enters the \textbf{exploitation} phase greedily by choosing the best possible action next. After each $\mu$ step, where $\mu$ is a hyperparameter, a random batch is sampled out of the \textbf{replay memory} to train the DQN.

Learning rate is set to $10^{-4}$.
Initial epsilon for exploration starts at 1.0 and drops to a minimum of 0.01 over 20000 timesteps.

We trained our models on Intel Xeon E5-2690 and Intel Xeon Gold 5122 for both manual and ODF sets of sub-sequences for x86 and AArch64 processor.
The number of time steps per iteration was set to 1005. It took around 16 hours to train the model on the CPU mentioned.
We study the improvements in code size and runtimes on Intel Xeon E5-2697 processor.
And, we study the improvements in code size in case of AArch64.
This is obtained by cross compiling LLVM to target Cortex-A72 processor.

\subsection{Results}
We evaluated the performance of our trained model by validating programs from MiBench~\cite{mibench} and SPEC-CPU (2006 and 2017) benchmarks~\cite{spec06:journals/sigarch/Henning06,spec17-Bucek:2018:SCN:3185768.3185771}. By choosing these set of benchmarks we cover both smaller and larger sized datasets. To check the performance of our trained model, for the above mentioned benchmarks, we apply the optimization sequence as suggested by the trained model. Then, we generate the binaries and compute the binary size for all the programs. We then compare it with the binary size generated with \texttt{Oz} optimization for the respective program.

In Table~\ref{tab:result_size}, we show the min, avg and max \% of size reduction for both x86 and AArch64 processor when we use sequences predicted by the model that is trained with manual and ODG sub-sequences. Negative value indicates that the binary size obtained by applying the predicted sequence is more than the binary size obtained using \texttt{Oz}. Min represents the \% of the minimum size improvement. Or, if it is negative, it represents the maximum size increased. Max represents the maximum size reduction, while avg represents average/mean size reduction for the respective benchmark.

\begin{table}[!htbp]
\renewcommand{\arraystretch}{1.5}
    \centering
    \caption{Percentage of min, mean and max size reduction wrt manual and ODG sequences for x86 and AArch64}
    \begin{tabular}{||p{0.08\textwidth}||p{0.041\textwidth}|p{0.035\textwidth}|p{0.035\textwidth}||p{0.041\textwidth}|p{0.035\textwidth}|p{0.035\textwidth}||}
        \hline
         \textbf{Benchmark} & \multicolumn{3}{c||}{\textbf{Manual sub-sequences}} & \multicolumn{3}{c||}{\textbf{ODG sub-sequences}} \\
         \hline
         & \textbf{Min} & \textbf{Avg} & \textbf{Max} & \textbf{Min} & \textbf{Avg} & \textbf{Max} \\
        \hline
        
        & \multicolumn{6}{c||}{x86} \\
        \hline
        \textbf{SPEC-2017} & -2.14 & 0.12 & 3.74 & -1.63	& 6.19 & 22.94 \\
        \hline
        \textbf{SPEC-2006} & -3.69 & -0.56 & 2.45  & -0.02 & 4.38 & 9.93  \\
        \hline
        \textbf{MiBench} & -4.82 &	-1.26 &	0.91 & -1.28 & 1.87 & 8.68 \\
        \hline
        \hline
        
        & \multicolumn{6}{c||}{AArch64} \\
        \hline
        \textbf{SPEC-2017} & -8.45 & 0.88 & 4.88 & -0.99 &	5.33 & 20.29 \\
        \hline
        \textbf{SPEC-2006} & -5.16 & 2.47 & 6.64  & -0.82 & 5.04 & 9.58 \\
        \hline
        \textbf{MiBench} & -9.43 &	-2.31 &	0.54 & -7.54 & 0.01 & 7.20 \\
        \hline
        \hline
    \end{tabular}
    \label{tab:result_size}
\end{table}

Table~\ref{tab:result_size} shows the performance, in terms of minimum, average and maximum reduction in code size, of the model when using manual sub-sequences and ODG sub-sequences when compared to \texttt{Oz}. 
We observe that using manual sub-sequences can achieve maximum size reduction of $3.74\%$ for SPEC-CPU 2017 and $2.45\%$ for SPEC-CPU 2006 for x86. 
Similarly, for AArch64 we obtain maximum size reduction of $4.88\%$ for SPEC-CPU 2017, and $6.64\%$ for SPEC-CPU 2006.

We also observe that for binary size reduction, the model with ODG sub-sequences performs better than the model with manual sub-sequences  for all the considered benchmarks on both x86 and AArch64 processors. With ODG sub-sequences, the average size reduction is $6.19\%$ and $4.38\%$ with a maximum of $22.94\%$ and $9.93\%$ size reduction for SPEC-CPU 2017 and SPEC-CPU 2006 respectively for x86. Likewise, the average size reduction for SPEC-CPU 2017 is $5.33\%$ with a maximum of $20.29\%$ size reduction for AArch64.
It can be seen that average size reduction with ODG sub-sequences is positive for all the benchmarks and a maximum of $6.19\%$ for SPEC-CPU 2017 for x86 processor is achieved. 
Although we achieve good maximum size reduction with manual sub-sequences, on an average it is not able to reduce binary sizes when compared to that of ODG sequences. 
In the sequences obtained from ODG, there are very few benchmarks where the binary size has increased in comparison with \texttt{Oz} optimization.

\begin{figure*}[h]
    \centering
    \subfloat[SPEC Runtimes]{{\includegraphics[scale=0.5]{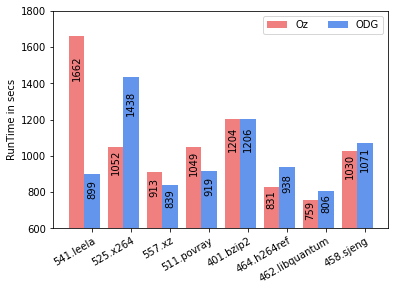} }}%
    \subfloat[SPEC Runtimes]{{\includegraphics[scale=0.5]{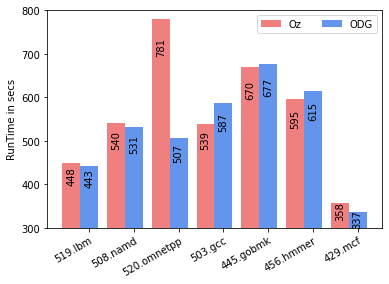} }}%
    \qquad
    \subfloat[SPEC code size]{{\includegraphics[scale=0.5]{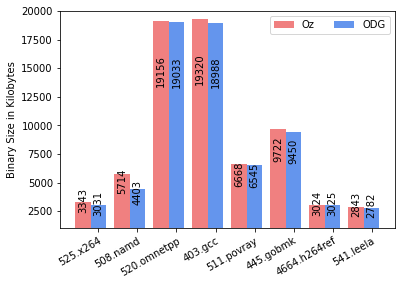} }}%
    \subfloat[SPEC code size]{{\includegraphics[scale=0.5]{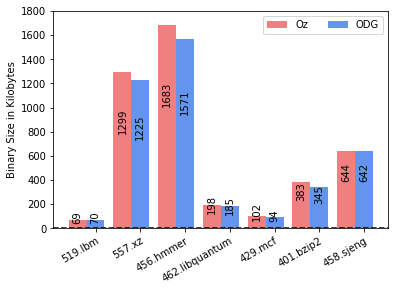} }}%
    
    \caption{Runtime and Binary size for benchmarks from SPEC-CPU suite (2006 and 2017) for \texttt{Oz} and ODG sub-sequences (lower is better)}
    \vspace{-\baselineskip}
    \label{fig:spec_runtime}
\end{figure*}

We also run the binaries and measure their execution time and compare with the execution time of \texttt{Oz}.
The detailed analysis of runtime and code size for the binaries generated by predicted sequences from ODG for SPEC-CPU 2017 and SPEC-CPU 2006 wrt x86 processor are shown in Fig.~\ref{fig:spec_runtime}.

\begin{table}[!htbp]
\renewcommand{\arraystretch}{1.5}
    \centering
    \caption{Percentage of improvement in execution time wrt manual and ODG sequences for x86}
    \begin{tabular}{||p{0.08\textwidth}||p{0.075\textwidth}||p{0.075\textwidth}||}
        \hline
         \textbf{Benchmark} & \multicolumn{1}{c||}{\textbf{Manual sub-sequences}} & \multicolumn{1}{c||}{\textbf{ODG sub-sequences}} \\
         \hline
        \hline
        \textbf{SPEC-2017}  & 7.33 & 11.99 \\
        \hline
        \textbf{SPEC-2006} & -4.68 & -4.19 \\
        \hline
        \textbf{MiBench} & 4.13 & 6.00 \\
        \hline
    \end{tabular}
    \label{tab:result_runtime}
\end{table}

Table~\ref{tab:result_runtime} shows the $\%$ decrease in execution time of binaries obtained by applying the sequences predicted by the model trained with manual or ODG sub-sequences. We can observe that using ODG sub-sequences reduces execution time by $11.99\%$ for SPEC-CPU 2017 and $6.00\%$ for MiBench

Fig.~\ref{fig:spec_runtime} shows the execution time improvement in seconds and binary size reduction in kilobytes for SPEC-CPU 2017 and SPEC-CPU 2006 benchmarks. 
With the ODG sub-sequences, the model predicts a sequence that leads to a reduction in execution time for \texttt{541.leela} by $45.91\%$ and \texttt{520.omnetpp} by $35.08\%$.
We achieve good performance improvement for most of the SPEC-CPU 2017 benchmarks, while execution time is increasing for most of the SPEC-CPU 2006 benchmarks.

In Fig.~\ref{fig:spec_runtime}(c) and \ref{fig:spec_runtime}(d), we show the percentage reduction in binary size for the sub-sequences generated by applying ODG sequences when compared to the binaries generated by applying \texttt{Oz} on SPEC-CPU 2017 and SPEC-CPU 2006 benchmarks. It can be seen that for almost all the benchmarks, our framework is able to reduce the binary size. However, for 519.lbm and 464.h264 there is a slight increase in binary size.

\begin{table}[!htbp]
    \centering
    \renewcommand{\arraystretch}{1.4}
    \caption{Some predicted sub-sequences}
    \begin{tabular}{||p{0.004\textwidth}|p{0.44\textwidth}||}
        \hline
        1 & 9\textrightarrow 31\textrightarrow 15\textrightarrow 6\textrightarrow 20\textrightarrow 14\textrightarrow 25\textrightarrow 5\textrightarrow 11\textrightarrow 27\textrightarrow 0\textrightarrow 10\textrightarrow 30\textrightarrow 24\textrightarrow 4 \\
2 & 15\textrightarrow 7\textrightarrow 23\textrightarrow 21\textrightarrow 30\textrightarrow 5\textrightarrow 14\textrightarrow 0\textrightarrow 25\textrightarrow 31\textrightarrow 6\textrightarrow 2\textrightarrow 27\textrightarrow 10\textrightarrow 11 \\
3 & 24\textrightarrow 30\textrightarrow 3\textrightarrow 31\textrightarrow 6\textrightarrow 28\textrightarrow 5\textrightarrow 17\textrightarrow 0\textrightarrow 18\textrightarrow 8\textrightarrow 2\textrightarrow 25\textrightarrow 22\textrightarrow 27 \\
4 & 14\textrightarrow 11\textrightarrow 7\textrightarrow 32\textrightarrow 30\textrightarrow 10\textrightarrow 23\textrightarrow 9\textrightarrow 12\textrightarrow 18\textrightarrow 2\textrightarrow 16\textrightarrow 13\textrightarrow 4\textrightarrow 0 \\
5 & 14\textrightarrow 11\textrightarrow 19\textrightarrow 10\textrightarrow 16\textrightarrow 32\textrightarrow 30\textrightarrow 9\textrightarrow 12\textrightarrow 23\textrightarrow 26\textrightarrow 13\textrightarrow 7\textrightarrow 17\textrightarrow 15 \\
        \hline
    \end{tabular}
    \label{tab:top5sequences}
\end{table}

In Table~\ref{tab:top5sequences}, we show 5 sub-sequences predicted by the model. The first sequence corresponding to x86 \texttt{508.namd} starts with initial passes from the \texttt{Oz} sequences such as \texttt{-simplifycfg}, \texttt{-sroa}, \texttt{-early-cse} and then has loop passes, \texttt{-mem2reg}, intermediate \texttt{Oz} passes such as \texttt{-globalopt}, \texttt{-globaldce} and ends with loop passes. 
The second sequence was predicted for \texttt{525.x264} on x86. It alternates between initial/intermediate \texttt{Oz} passes and loop passes. Whereas, the third sequence starts with loop passes, initial passes from \texttt{Oz}, \texttt{-mem2reg} and then has intermediate \texttt{Oz} passes and ends with loop passes. This sequence was predicted for \texttt{susan} benchmark from MiBench in x86.
The fourth sequence has intermediate \texttt{Oz} passes, initial \texttt{Oz} passes, \texttt{-mem2reg}, loop passes and initial/intermediate \texttt{Oz} passes in that order.
The last two sequences correspond to \texttt{508.namd} and \texttt{511.povray} benchmarks on compiling for AArch64.

As it can be observed, the predicted ODG sequences contain a mix of passes from initial, intermediate, ending pass sub-sequences and loop passes of \texttt{Oz}. 
Such combination of sub-sequences are not present in the \texttt{Oz} sequence.
This shows that our framework can exploit the performance by applying the new sub-sequences which otherwise would have not been applied.
It can also be seen that different sub-sequences are predicted for different sources, that in turn improves the performance.

%% file: relatedWorks.tex
\section{Related Works}
\label{sec:relatedworks}

In recent days, several approaches that use machine learning to get better heuristics for many compiler optimization problems like vectorization~\cite{Stock2012}~\cite{Neurovectorizer-HajAli-2020}, inlining~\cite{Simon-Inlining-10.1109/CGO.2013.6495004}, unrolling~\cite{Stephenson2005}, etc. have been proposed and are widely accepted. 

Several approaches have been proposed to use machine learning techniques to solve the phase ordering problem.
These works range from using genetic algorithms~\cite{Cooper99}, \cite{Kulkarni}, clustering techniques~\cite{MiCOMP}, and Bayesian networks~\cite{bayesian}. 
Also, several such approaches have been proposed to optimize for either execution time~\cite{MiCOMP}, \cite{PuriniS} or code size~\cite{Cooper02},~\cite{YACOS}.
To our understanding, ours is the first framework that attempts to predict optimization sequences that jointly optimizes both execution time and binary size .
There have been other works including those that optimize compile time~\cite{Kulkarni} and power consumption~\cite{energyconsumption}. 

Most of these works are based on \textit{expert defined features}. 
However, the process of defining the complete set of features is cumbersome and needs domain compiler engineer expertise. Hence, of late, deep learning is gaining popularity with its ability to process large scale of data, and as it can relieve the expert to pick the right set of features. 
The deep learning based approaches do not require feature engineering by the expert compiler engineer.

Several approaches have been proposed to represent source code as input to deep learning methods.
Feature based approaches represent programs as collection of hand picked features like number of basic blocks, number of branches, etc. 
Milepost~\cite{Fursin2008MILEPOSTGM} is one such work that uses program features to learn the better optimization sequences and the parameters for the optimization algorithms.

On the other hand, there exist approaches that represent source codes as \textit{distributed vector}, where the meaning of the code is \textit{distributed} across various components of the high-dimensional vector.
Methods like code2vec~\cite{alon2019code2vec}, inst2vec~\cite{ncc} and IR2Vec~\cite{IR2Vec} fall under this category.
These approaches use different abstractions of the program for forming representations.
code2vec uses the paths of Abstract Syntax Trees, while inst2vec and IR2Vec use the Intermediate Representation of the programs to form language independent representations.
These approaches come up with representations that are independent of the underlying applications.

Several approaches that use DNNs on reinforcement learning setup have been proposed to solve the phase ordering problem~\cite{Mammadli}, \cite{AutoPhase}.
Modelling phase ordering for compiler optimizations as a reinforcement learning fits naturally, as the number of possible combinations of the optimizations are too high even on fixing the sequence length. For a sequence of length $n$ and for the number of passes $p$, a total of $p^n$ combinations are feasible. 
Enumerating them to create a dataset to train on is intractable.
Hence our framework also models the phase ordering problem as a reinforcement learning problem.

Different approaches have been proposed to model the output space differently. 
A seminal work by Cooper et al.~\cite{Cooper99} proposed an ordering on 10 transformation passes to optimize for code size.
Another important work is by Kulkarni et al.~\cite{Kulkarni} that shows that such a space of optimization sequences is too big to enumerate, and devise an exhaustive search technique by which they analyze this space to automatically calculate relationships between different phases.

The recent approaches model the output space as sub-sequences, where the sub-sequences are often grouped manually by following certain heuristics.
Ashori et al. proposed a framework, MiCOMP~\cite{MiCOMP} where the passes in \texttt{O3} sequence are clustered to form groups.
These clusters are then treated as sub-sequences. A predictive model learns the ordering of these sub-sequences to achieve better execution time.

Jain et al.~\cite{Jain2019AnAO} analyze the effects of optimizations in \texttt{Oz} sequence. They divide the \texttt{Oz} sequence into a set of \textit{logical groups} and analyze them to study their contributions for the code size reduction.

Silva et al~\cite{silva} in a similar spirit designed an approach to arrive at the reduced set of optimization passes that perform better than \texttt{Os}. They come up with a covering set of optimization passes by exploring the space of optimizations using genetic algorithms.

Closest to our work is that of Mammadli et al.~\cite{Mammadli}
They propose a framework that uses reinforcement learning to come up with optimization sequences that have better execution time characteristics when compared to \texttt{O3}. 
They model action spaces by dividing the optimizations in \texttt{O3} into sub-sequences.
These sub-sequences are divided into different levels of abstraction, where an action at a higher level has more number of passes and parameters, while the lowest level has a single pass with parameters.  For a given program, the predicted sequence has a fixed length and usually has a repetitions of the same action.
We on the other hand, design a framework to achieve better code size characteristics while preserving, if not improving the run time, when compared to \texttt{Oz} by using a double DQN model.
Also, we propose two different output spaces as actions - a sub-sequence based output space derived from \texttt{Oz} in the similar spirit of Mammadli et al; sub-sequences obtained by traversing over the \textit{Oz dependence graph} (ODG).
Our methodology is also different, we use the novel IR2Vec embeddings~\cite{IR2Vec} that can use the LLVM-IR representation for encoding programs. 

%% file: conclusion.tex
\section{Conclusions}
\label{sec:conclusion}

We proposed a reinforcement learning based framework to solve the compiler phase ordering problem, that results in improving \textbf{both} code size and execution time.
We model an action space using two approaches, one based on manual sub-sequences, and other by creating a graph out of \texttt{Oz} optimization sequence and traversing it.
Using only minimal training, we were able to obtain improvements in terms of code size and execution time over \texttt{Oz} optimization flag of LLVM-10.
Our model uses static rewards calculated at the compile time relying on LLVM-MCA for the runtime, and the binary size to model code size characteristics.
We show the code size improvements on x86 and AArch64 architectures, and execution time on x86 using standard benchmarks.

Our approach can be extened to \texttt{O3} or other optimizations by constructing the corresponding pass dependence graphs for optimizing the performance or size.
In future, we plan to extend this framework to support predicting the parameters of the optimizations (like unroll factors and vector factors) along with the sequence.
We have integrated our framework with LLVM 10. The source code and other relevant artifacts are available at \url{https://compilers.cse.iith.ac.in/projects/posetrl/}.